\documentstyle[12pt,aasms4]{article}
\def\be{\begin{equation}}
\def\en{\end{equation}}
 
\begin{document}
\title{On the role of shock waves in galaxy cluster evolution}
\author{Vicent Quilis
, Jos\'e M$^{\underline{\mbox{a}}}$. Ib\'a\~nez
\and Diego S\'aez}
\affil{Departament d'Astronomia i
Astrof\'{\i}sica, Universitat de Val\`encia,
E-46100 Burjassot, Val\`encia, Spain}
\authoremail{vicente.quilis@uv.es}

\begin{abstract}

Numerical simulations of galaxy clusters including two 
species -- baryonic gas and dark matter particles --
are presented. 
Cold Dark Matter spectrum, Gaussian statistics and
flat universe are assumed. 
The dark matter component is evolved numerically 
by means of a standard {\it particle mesh} method. 
The evolution of the baryonic component has been 
studied numerically by using 
a multidimensional (3D) hydrodynamical code based on
{\it modern high resolution shock capturing} techniques. 
These techniques are specially designed for treating accurately
complex flows in which shocks appear and interact.
With this picture, the role of
shock waves in the formation and evolution of rich galaxy clusters 
is analyzed. Our results display two well differenced morphologies of the 
shocked baryonic matter: filamentary at early epochs and quasi-spherical 
at low redshifts.

\end{abstract}


\section{Introduction}

Galaxy clusters are the largest systems gravitationally bounded in
the Universe. Their study has been a fashion topic in Cosmology
since last years. Work on topics related with
galaxy clusters is worthly to:
i) understand
the formation, evolution, dynamics and morphology of these 
systems, ii) learn on the physical processes 
involved in them, and iii)
find out some information concerning with 
fundamental parameters in Cosmology as density parameter ($\Omega$)
, Hubble's constant ($H$), and the spectrum of the primordial
density field. 

During last years technical improvements
have produced huge quantity of data about galaxy clusters. 
Let us mention the 
new galaxy surveys (Guzzo 1996, and references therein) and the
extensive observation in X-rays using
satellites as ROSAT or ASCA. These huge volume of 
data strongly motivates a lot of theoretical work
trying to explain the observational results.

From the theoretical point of view, numerical simulations are the best tools
to understand physics involved in galaxy clusters.
At the beginning, numerical simulations of galaxy clusters where performed 
using N-body techniques. Since then, they have been extensively used 
and  have produced important results ( see, e.g, 
Efstathiou et al. 1985, 
Bertschinger \& Gelb 1991, Xu 1995). Next step in the full 
description of galaxy clusters 
was to introduce in the picture a baryonic component. The numerical methods 
developed in order to deal with baryonic matter were more sofisticated and 
expensive in computational resources. 
As a consequence, it was not possible to carry out numerical simulations 
with two species (dark matter and baryonic gas) until late eighties.

Cosmological hydrodynamic codes have been usually classified in two
main categories: a) the so-called Lagrangian methods, like the
{\it Smoothed Particles Hydrodynamics} (SPH) or ulterior extensions
 based on them, and b) Eulerian codes.

SPH methods were first proposed by 
Gingold \& Monaghan (1977), and  Lucy (1977). 
Among the best features of this technique, it
should be pointed out its high resolution in dense regions. This 
property is directly derived from its Lagrangian character.
The first implementations of SPH techniques 
had some weak points:
i) The low density regions were badly described due to
the Lagrangian character of the method. 
ii) Discontinuities and strong gradients were poorly 
solved and an important diffusion was introduced.
iii) They were not conservative. 
Nevertheless, these previous problems were overcome in the 
modern implementations of these techniques.
Improved SPH techniques
have been widely developed for cosmological applications 
(see, e.g.,
Evrard 1988, Hernquist \& Katz 1989, Navarro \& White 1993, Gnedin 1995).

Numerical cosmological codes using an Eulerian approach to study 
baryonic gas inside galaxy clusters have been also developed.
Some of these hydro-codes use {\it
artificial viscosity} in order to deal with shock waves 
(Cen 1992, Anninos et al. 1994).
These techniques require a good calibration of the free parameters
which are introduced by hand and 
state some numerical problems. 
Recently, a new family of finite difference methods, which
use Eulerian approaches and avoid artificial viscosity, has been developed
in numerical Cosmology. They are the 
so-called {\it high resolution shock capturing 
methods} (HRSC), the modern extensions of the original 
Godunov's idea (1959).
According to the Riemann solver and the procedure in order
to achieve spatial accuracy,  we can distinguish
three groups: 1) the ones following Harten's scheme (1983), like 
Ryu et al. (1993), 2)  those using 
the analytical solution of the Riemann problem for 
the Newtonian dynamics of ideal gases
and the PPM scheme described by Collela \& Woodward (1984)
, like Bryan et al. (1994), and 3) the codes using 
Roe's Riemann solver (Roe 1981) plus the MUSCL or PPM
cell reconstruction, like
in Quilis et al. (1996).
In this last reference, the code used
in present paper is described and tested appropriately.

An exhaustive comparison among all these kinds of cosmological
hydrodynamic codes can be found in 
Kang et al. (1994).

Due to the Eulerian character of our code, it does not show
--in dense regions-- a resolution as good as the Lagrangian
ones, and it requires more computational resources. 
However, HRSC schemes --by construction-- have 
excellent properties in order to deal with shocks, 
discontinuities, and strong gradients. 
HRSC techniques typically solve shocks in two cells. 
Due to their intrinsic properties, the detection of shocks
is independent on the number of cells used in the simulations.
It should be pointed out that this last property is really important
when three-dimensional simulations are carried out. In these
simulations the size of the grid is a stringent constraint
due to its high cost in computational resources. 
Moreover, these methods are  conservative by construction, 
that is , quantities which should be 
physically conserved are numerically conserved up to 
the order of the method.  It should be also noticed that
these methods show good results in 
extreme low density regions (Einfeldt et al. 1991).

As it has been pointed out by several authors, the role of
shock waves can be extremely important in order to understand
the heating processes in the intracluster medium (ICM). In this 
paper we are interested in understanding and quantifying the role
of shocks. In order to do that it is crucial to use numerical 
codes able to manage with complex flows. 
Yes, indeed, one of the important features of HRSC techniques 
is just to treat numerically shocks and strong discontinuities
giving sharp profiles (in a few numerical cells, as we have
mentioned above) independently of the size of the grid.
Hence, formation, evolution, and interaction of shocks in 3D flows
can be analyzed accurately with HRSC schemes, and, consequently,
their use is absolutely 
justified in order to study shocks and their consequences
on the ICM's dynamics.   

Hereafter, $t$ stands for the cosmological time, $t_0$ is the age of the
Universe, $a(t)$ is the scale factor of a flat background. 
Function $\dot{a}/a$ is denoted by $H$, where the dot stands for
the derivative of $a$ with respect to the cosmological time.
Hubble constant is the present value of
$H$; its value in units of $100 \ Km \ s^{-1} \ Mpc^{-1}$ is 
the reduced Hubble constant $h$. In our computations we have
assumed $h=0.5\, $. Velocities
are given in units of the speed of light. 
Baryonic, dark matter, and 
background mass density are denoted by $\rho_{_b}$, $\rho_{_{DM}}$, and 
$\rho_{_{B}}$, respectively. The density contrast
is $\delta_b=(\rho_b-\rho_{_{B}})/\rho_{_{B}}$ for baryonic matter, 
analogously is defined $\delta_{_{DM}}$ for dark matter.

The plan of this paper is as follows: In Section 2, our numerical cluster
model is described. In Section 3, the results of the simulations are 
analyzed. Finally, a
general discussion is presented in Section 4.

\section{Cluster Model}

\subsection{Initial conditions}
Initial conditions are given at redshift $z=100$. We have
considered a flat universe $(\Omega_o=1)$ and a CDM scenario. The
density profile is obtained by using the power
spectrum:
\be
\label{spec}
|\delta_k|^2=\frac{{\cal A}k}{(1+\beta k + \omega k^{3\over 2} +
\gamma k^2)^2}
\en
where $\beta=1.7(\Omega_o h^2)^{-1} \, Mpc$, $
\omega=9.0(\Omega_o h^2)^{-1.5} \, Mpc^{1.5}$, $\gamma=(\Omega_o h^2)^{-2} 
\, Mpc^2$, and the constant ${\cal A}$ (normalization) has been fitted by
using $\sigma_8=0.63$, where 
$\sigma_8\equiv\langle ({\delta\rho \over \rho})^2 \rangle^{1\over2}$ 
on the 8 $h^{-1}\, Mpc$ scale.
Assuming that the density field is Gaussian, 
realizations in the position space can be obtained from Eq.(\ref{spec}). 

Hoffman and Ribak (1991) described a method to
produce realizations of Gaussian fields under some constraints.
This procedure was extended by van den Weygaert and Bertschinger (1996).
We have used this powerful method in order
to obtain the initial conditions for our cluster. In particular, we 
have assumed
three constraints: 
1) the cluster peak is located just at the center of the box,
its amplitude corresponds to  
the maximum density contrast in our initial box (64 Mpc per 
edge in comoving coordinates), 
2) the mass enclosed inside a Gaussian ball,
$M=(2\pi)^{(3/2)}\rho_BR_f^3$ with $R_f=4 \,h^{-1}$ 
comoving Mpc, is  $M=6\times 10^{14}\, M_{\odot}$
--centered at the peak position-- , and 
3) the amplitude of the initial density 
contrast --cluster peak-- is $3\sigma$ after smoothing on 
scales $\sim 4 h^{-1} Mpc$ using a Gaussian window function.

The 3D Zeldovich's approximation is used to evolve from 
the above initial conditions up to $z=30$. At this redshift, 
a comoving box of 20 Mpc per edge is extracted (zoom)
from the initial box. The maximum
of the density contrast 
is located at the center of the new box.

The density and the velocity profiles of the initial perturbation
are known after performing the zoom at $z=30$. Then, 
we define the composition of this perturbation, namely, we fix the
ratio between the 
dark matter (DM) and baryonic densities. It is considered 
that the 90 \% of the matter is dark and the remaining is baryonic. 
The DM component is
discretized into particles of the same mass (the mass of the
particles depend upon the resolution of each simulation). 
The baryonic component is considered as an ideal fluid 
of adiabatic exponent $\gamma$, fully ionized, 
with mass abundances of 75\% Hydrogen and
25\% Helium, which represents the intracluster medium
(ICM). We assume that the initial velocities of both components are 
identical.
The initial homogeneous
temperature is set up assuming that at $z=30$ the temperature is given
by $\displaystyle{T_{30}=T_{200}\left(\frac{\rho_B(z=30)}
{\rho_B(z=200)}\right)^{\gamma-1}}$,
where $T_{200}=2.73(1+z)$ with $z=200$,
 and $\gamma$ is the adiabatic exponent (see , e.g., Anninos \&
 Norman 1996).

The above initial conditions give the seed for the formation of
a large Abell cluster. 

Some comments about the boundary conditions are deserved. 
The baryonic and DM have low densities at the faces of
the numerical box. Moreover,  DM particles
are not allowed to escape out the box. 
In order to do that, when one particle crosses some
of the faces of the computational box, it is introduced through the 
symmetric face with the same velocity. Thus, 
a reasonable approximation to periodic boundary conditions
is considered.

\subsection{Nonlinear evolution}

The nonlinear evolution of the baryonic component is described
by the hydrodynamics equations (Peebles 1980):
\begin{equation}
\label{hydro1}
\frac{\partial \delta_{_b}}{\partial t} + \frac{1}{a}\nabla\cdot 
(1+ \delta_{_b}){\vec v} = 0
\end{equation}
\begin{equation}
\frac{\partial {\vec v}}{\partial t} + \frac{1}{a}
({\vec v}\,\cdot\, \nabla){\vec v}
+ H{\vec v} = - \frac{1}{\rho_{_b} a}\nabla p - \frac{1}{a} \nabla \phi
\end{equation}
\begin{equation}
\label{hydro2}
\frac{\partial E}{\partial t} + {1\over a} \nabla\cdot[(E + p) {\vec v}] =
-3H(E+p) -H\rho_{_b} {\vec v}^2 - \frac{ \rho_{_b} {\vec v}}{a} \nabla \phi
- \Lambda 
\end{equation}
and the evolution of the DM particles obeys the particle equations:
\be
\label{part1}
\frac{d{\vec x}}{dt}= \frac{\vec v}{a}
\en
\be
\label{part2}
\frac{d{\vec v}}{dt}= - \frac{\vec{\nabla \phi}}{a} - H{\vec v}
\en 
\noindent
where ${\vec x}$, ${\vec v}=a(t)\frac{d{\vec x}}{dt}= (v_x, v_y, v_z)$, $E$
and $\phi(t,{\vec x})$ are,
respectively, the Eulerian coordinates, the peculiar velocity,
the total energy $E= \rho_{_b} \epsilon + {1\over 2} \rho_{_b} v^2$
($v^2 = v_x^2 + v_y^2 + v_z^2$),
 and the peculiar
Newtonian gravitational potential. 
The pressure and the internal energy 
per unit mass are denoted, respectively, as 
$p$ and $\epsilon$. 
The above system of hydrodynamical equations is closed with 
the equation of state of an ideal gas $p=(\gamma -1 )\rho_{_b}\epsilon$ and
$\gamma=5/3$ (monoatomic gas).

In some applications described in the present paper two cooling
processes are taken into account, Compton cooling ($\Lambda^C$) and
thermal Bremsstrahlung ($\Lambda^{Br}$). The well known 
expressions for these processes are (see, for instance, 
Umemura \& Ikeuchi (1984)):
\be
\Lambda^C=5.4\times10^{-36}(1+z)^4 n_e T \, \, \, \, 
(erg cm^{-3}s^{-1})
\en
and
\be 
\Lambda^{Br}=1.8\times10^{-27}n_eT^{1\over2}
(n_{HII} + 4n_{HeIII}) \, \, \, \, (erg cm^{-3}s^{-1})
\en
where $n_e,n_{HII}$, and $n_{HeIII}$ are the number density
of electrons, protons, and helium nuclei, respectively.
The cooling effects are introduced in the source  
term $\Lambda=\Lambda^C+\Lambda^{Br}$ in Eq.(\ref{hydro2}).

Both components, gas and particles, are gravitationally 
coupled through Poisson's equation:
\begin{equation}
\label{poisson}
\nabla^2\phi = \frac{3}{2} H^2 a^2 \delta_{T},
\end{equation}
where $\delta_{T} = (\rho_{_b}+\rho_{_{DM}} - \rho_{_B})/ \rho_{_B}$
is the total density contrast.

The hydrodynamical equations (\ref{hydro1}-\ref{hydro2})
define a {\it hyperbolic system of conservations laws} (with sources).
This important property allows us to apply a new powerful 
family of numerical methods (HRSC) in order to solve this system. 
The main features of these methods are summarized: 
1) the quantities which should be physcially conserved, 
are numerically conserved  by
construction, due to the fact that the algorithm is written in 
conservative form                              
, 2) these methods are of high order in the smooth regions of the
flow, 3) shocks are sharply solved in typically two 
numerical cells,  independently of the number of cells used in the
simulations
, 4) numerical artifacts like 
{\it artificial viscosity} are avoided, 5) these methods are free of 
spurious oscillations around shocks, and 
6) they are able to handle strong gradients and discontinuities.
In Quilis et al. (1996) we took advantage of the hyperbolicity property
of this system in order to design a multidimensional hydrodynamic code.
A full description of this code and a set of tests proving its 
performance were presented in this reference. The above 
method is used in present
paper in order to study the evolution of the baryonic gas.

The evolution of DM particles described by Eqs. (\ref{part1}-\ref{part2}) 
is studied through a standard Particle Mesh (PM) method 
(Hockney \& Eastwood 1988). This method has some important features
from the point of view of our applications: 1) the PM methods 
are widely extended and well known, 2) they are easily  
programmable and can be directly coupled with a hydro-code,
and 3) their CPU cost is quite low in comparison with others 
particle type methods. This last property is crucial when the 
particle code must be coupled with a hydro-code like the one
described above where the computational effort is severe. 

Poisson's equation (\ref{poisson}) is solved by using Fast
Fourier Transform (FFT) methods (Press et al. 1987). 
In order to recover a continuous density contrast field 
for dark matter component, we use a standard cell-in-cloud (CIC)
scheme (Hockney \& Eastwood 1988) at each time step.

The time step is computed following Quilis et al. (1996).
We have considered Courant's and dynamical characteristic  times. 
These times must be corrected in order to ensure the 
numerical stability of the algorithm. Typical values of
these correction factors, known as CFL, are CFL$_{Courant}=0.6$
and CFL$_{dynamical}\sim10^{-2}$.
Besides the characteristic times mentioned before, cooling processes
introduce a new characteristic time. Nevertheless as it is known 
(Tsai et al. 1994),            
for the scales considered in this paper, the characteristic cooling
time is  larger than the dynamical one and, consequently, cooling 
would not expect to be an effective process.

In order to display the conservation properties of the whole code -- 
hydro, particles and the Poisson solver --  the
global energy conservation is investigated in the Appendix A.
Some comments on the
numerical spatial resolution of our simulations are also 
discussed in Appendix B.

\section{Results}

In this Section we are going to discuss the results obtained
running our code in two cases: i) {\it without
cooling} processes (denoted by A), and ii) {\it including
cooling} effects (denoted by B). 

Numerical simulations have been carried out using 
128$^3$ DM particles and 128$^3$ baryonic cells. 
The physical comoving box size is 20 Mpc per edge.
Since we are interested in the study of ICM and not 
concerned with smaller scales like galaxies theirselves,
the choice of 128$^3$ DM particles and 128$^3$ baryonic cells
seems adequate. 
The resolution of the 
hydrodynamical code is $\sim 0.15$ comoving $Mpc$, however the PM
code resolutions is two cells. 
Hence, the minimal resolution of our simulations will
be around $\sim 0.3$ comoving $Mpc$. 
Taking into account
the properties of HRSC techniques,
the grid used in present paper seems a reasonable compromise
between the resolution needed in order to describe physical 
processes and our current computational limitations.

The CPU cost of our numerical simulations is 
183.2 s(25.1 s) per time step 
(with $128^3$ numerical cells, and $128^3$ particles)
in a Hewlett-Packard J280 (Silicon Graphics
Origin 2000 with 8 processors for parallel version).

\subsection{Cluster evolution}

In the standard CDM scenario, clusters of galaxies grow by
merging from smaller collapsed structures. This behaviour
can be tested through N-body simulations .
Figure 1 shows the DM particle distribution
for our A simulation at six epochs. 
DM dynamics seems to be not much affected by 
gas dynamics and cooling effects. By this reason
the results from B 
simulations are not shown. In Figure 1,
32$^3$ particles have been plotted. The edges of the boxes are 
20 comoving $Mpc$ length.

%%%%%%%%%%%%%%%%%%%%%%%%%%%
% figura 1 , slices de materia oscura
%%%%%%%%%%%%%%%%%%%%%%%%%%%%%%%%%%%%%

The implementation of a baryonic gas complicates the description
of the cluster. This is a component with pressure and with
a different dynamics. This new component could undergo compression,
expansion, heating and  cooling. Non-adiabatic processes like shock 
waves and the cooling -- due to Bremsstrahlung  
and Compton -- could also be present. These deviations from 
adiabaticity are very 
important in order to understand the mechanism which provides
an extra energy to ICM. Figure 2 shows ICM density contrast 
colour contours for A simulation. 
The slices are 20 (0.15) comoving Mpc length (depth).
Four times are 
displayed. The evolutionary picture starts with 
collapsing processes of small substructures.
These substructures undergo a merging process towards 
a well defined central object. This picture is valid for 
both simulations, A and B. 
Since no significant
differences can be found, 
for the sake of briefness only A results have been displayed.
These results supports the well known fact that 
for the scales considered in this paper, the cooling is 
not a relevant process (see ,e.g, Tsai et al. 1994).

%%%%%%%%%%%%%%%%%%%%%%%%%%%%%%%%%%%%%%
% Figure 2, slices de materia barionica
%%%%%%%%%%%%%%%%%%%%%%%%%%%%%%%%%%%%%%

A good description of the cluster evolution is given by
the profiles, at different epochs, of the main quantities averaged 
in spherical shells. These shells are located at
several radii from the cluster center.
This center is defined  as the minimum of the 
total gravitational potential. Each shell has a fixed comoving 
width. An average value inside the shell can be estimated for each 
variable. Figure 3 plots the average density contrast for 
baryonic gas (top panel) and 
DM (bottom panel) versus
radial distance to the cluster center in comoving Mpc. In 
each panel several curves are shown corresponding to  different
times. As no significant differences arise between A and B 
simulations, only case A is presented.
It should be pointed out the resemblance of the profiles plotted
here and the ones shown by Navarro et al. (1995). As it was
explained in detail in this reference, for the spatial 
resolution considered in this paper, both components evolve 
keeping the original ratios between their densities.

%%%%%%%%%%%%%%%%%%%%%%%%%%%%%%%%%%%%%%%%
% figura 3 , promedio radiales para delta
%%%%%%%%%%%%%%%%%%%%%%%%%%%%%%%%%%%%%%%%

The gross properties of the cluster dynamics 
can be fairly inferred studying 
average radial velocities of DM and baryonic gas. These 
averages are computed in a similar way to  the density
contrast. Figure 4 plots the average radial velocities, in units
of $c$, for baryonic gas (top panel)  and
DM particles (bottom panel).
As in Figure 3, only A results are shown.

%%%%%%%%%%%%%%%%%%%%%%%%%%%%%%%%%%%%%%
% figure 4 , promedio en velocidades
%%%%%%%%%%%%%%%%%%%%%%%%%%%%%%%%%%%%%%

From Figs. 3 and 4 the following stages in the evolution could be
roughly distinguished. First, an infalling phase until $z\sim 2$, 
characterized by a typical profile in velocity (gas and 
DM) and increasing in density contrast (both components). 
Let us point out that the profile, during the infalling epoch,
of the baryonic component remind us the typical one in any
collapsing object, the minimum of which separates a subsonic
homologously part (the inner one) from the supersonic freely
falling part (the outer one).
In a second
phase, gas velocities change their signs in the inner part   while
outer shells go on falling. Gas density contrast slightly decreases 
due to the gas outflow. 
Nevertheless, this effect is subdominant, and the ratios between
the averaged density profiles of gas and DM undergo small variations.
DM particles are not affected by this dynamics and 
exhibit some tendency to reach a dynamical equilibrium.    

\subsection{Looking for shocks}

In the present Section we are specially interested in 
giving a set of criteria allowing us to identify those
regions where shocks form and evolve.
In the next Section will analyze the energy released to the
ICM by these shocks.

From the top panel of Figure 4 some strong 
evidences of the existence of a quasi-spherical
shock are pointed out.
The velocity profiles show how the evolution 
begins with a typical infalling profile. This velocity
is strongly correlated with the growing of the 
baryonic density contrast, for the same time, in top
panel of Figure 3. As the density rises , the pressure 
becomes more important. At $z\sim 2$, the velocities at the
inner part of the cluster change their sign, and the
gas moves out. This is an evidence of a quasi-spherical 
shock moving outwards from the core cluster. 

Following Anninos and Norman (1996) we study the entropy
$s=ln(T/\rho^{\gamma -1}_{_{b}})$. In an adiabatic process this
quantity should keep constant. Variations in
$s$ are evidence of  non-adiabatic processes. In particular shock 
waves strongly influence the variations in the entropy. 
The maximum in the entropy could identify the position of
shocks.
Figure 5 plots the average entropy in spherical shells for 
A case at several times. 

%%%%%%%%%%%%%%%%%%%%%%%%%%%%%%%%%%%%%%%%%
% figure 5, promedios en la entropia
%%%%%%%%%%%%%%%%%%%%%%%%%%%%%%%%%%%%%%%%%

The average entropy profiles
evidence the existence of a quasi-spherical
shock travelling outwards from the cluster center.
The formation and evolution of the local maxima in
the average entropy profiles can be directly correlated 
with the changes in the velocity profiles for baryonic
gas (top panel Figure 4). 
The averaging procedure in shells is itself an important source of
numerical diffusion. When the shock deviates from sphericity or 
there are a set of smaller and irregular shocks, then
the average in spherical shells does not produce two
regions separated by a sharp jump in the entropy
profile.
In Figure 5 two regions are clearly differenced. The 
pre-shock region has a roughly constant entropy, 
while the post-shock shows a decreasing  inward  
entropy with values higher than those of pre-shock
region. 

The above analysis evidences the existence of, at least, one 
quasi-spherical shock.
Nevertheless, it is not possible to know if there is a unique quasi-spherical
shock, or there are some shocks at several scales which in 
average give the global shock in Figure 5. 
From this Figure, 
no ideas about the scales and form of 
the shocks can be obtained. 
In order to clarify
this point, Figure 6 shows slices for the
divergence of baryonic velocity, entropy, temperature and pressure.
These slices are centered at the point where the gas density is 
maximum, and are projected along z-axis. Size and depth of the 
slices are the usual through this paper (see Fig. 2).

%%%%%%%%%%%%%%%%%%%%%%%%%%%%%%%%%%%%%%%
% figure 6 ,slices for div,entropy....
%%%%%%%%%%%%%%%%%%%%%%%%%%%%%%%%%%%%%%%

As it has been mentioned above, entropy, pressure and temperature
exhibit a well known behaviour in shocks. Nevertheless, 
some comments on the usefulness of studying the 
divergence of baryonic velocity are worthwhile.
If we assume that, at a given time, the density contrast $\delta_b$ is 
homogeneous, and that 
the source terms in Eq. (\ref{hydro1}) are negligible, then, the 
time variation of the density contrast is related with 
the divergence of the velocity through the following 
equation,
\be
\frac{\partial ln\delta_{_b}}{\partial t}=-{1\over a}\nabla \cdot \vec v
\en
\noindent
This equation points out two different regimes: 1) 
compression, i.e., $\nabla \cdot \vec v < 0$, and  2) rarefaction, i.e.,
$\nabla \cdot \vec v > 0$. Therefore, the changes in the sign of
the divergence of the velocity allow us to distinguish between 
compressed and rarefied regions. This discussion can always be applied 
locally, and it gives a roughly description of the 
dynamical state of the system.

In our simulations,
neighbouring regions might be separated by a shock
if  simultaneously, $\nabla \vec{v}$ exhibits large changes 
, the entropy increases clearly, and  the temperature and pressure 
show strong gradients.  
The features of the slices for these quantities, allow us to have a good 
description of the shock scales and their morphology.

Strong evidences of a quasi-spherical shock travelling outwards
appear in all the quantities plotted in Figure 6.
A thin shell, where the divergence of baryonic velocity and temperatures
change strongly and the entropy reaches their maxima values,
is clearly appreciable.
Pressure exhibits a tiny jump in the same region, but, unfortunately, 
due to the extreme variations in pressure from the center to the outer 
regions, the colour scale is not able to reproduce this pressure jump.
 
Even more, it is possible to define some criteria which could
label a cell which is involved in a shock. Following the 
criteria used by Colella \& Woodward (1984) to detect shocks,
we generalize them for a multidimensional case. A numerical cell, $(i,j,k)$,
is labelled as involved within a shock if the following conditions are
satisfied:
\begin{eqnarray} 
\label{sh1}
& &\frac{|p_{i+l,j+m,k+n}-p_{i-l,j-m,k-n}|}{p_{i,j,k}} > \beta_1
\, \, \, \, \, l,m,n=0,1 \\
\label{sh2}
&max&\left(\frac{|v_{i+l,j+m,k+n}^q-v_{i-l,j-m,k-n}^q|}{v_{i,j,k}^q}\right) 
> \beta_2
\, \, \,  l,m,n=0,1 \, \,and \, \, q=x,y,z\\
\label{sh3}
& &\frac{|\rho_{b_{i+l,j+m,k+n}}-\rho_{b_{i-l,j-m,k-n}}|}
{\rho_{b_{i,j,k}}} > \beta_3
\, \, \, \, \, l,m,n=0,1 
\end{eqnarray}
\noindent
where $\beta_1,\beta_2$ and $\beta_3$ are three parameters to be fixed. 
Let us comment the above criteria to identify a cell as
a shocked one. As it is well known in 1D, across a shock, quantities 
such as 
$p,v$, and $\rho$ have a jump discontinuity. The lack of numerical
resolution may produce misleading results  without a fine
tuning of $\beta$ parameters. 
For example, in the case of low values of $\beta_1$, 
cells where there are only strong gradients could be 
identified as shocked cells.
These problems can be extended to the other conditions.
Some special attention deserves strong rarefactions. Since regions
where the gas is being fast rarefied could satisfy the above conditions
even with high values for $\beta$ parameters, one new condition must
be considered in order to avoid identifying strong rarefactions as 
shocks. This condition takes advantage from the fact that the 
velocity in the post-shock region is always greater than that of the 
pre-shock region (Colella \& Woodward 1984), i.e.,
\begin{eqnarray}
\label{sh4}
v^x_{i-1,j,k} > v^x_{i+1,j,k} \,\,\,\,\, or \,\,\,\,\,            
v^y_{i,j-1,k} > v^y_{i,j+1,k} \,\,\,\,\, or \,\,\,\,\,
v^z_{i,j,k-1} &>& v^z_{i,j,k+1} 
\end{eqnarray}

Some experimentation has been carried out so as to get the values  
for the parameters $\beta_1,\beta_2$, and $\beta_3$, and also
to check their performance.  As we have seen above, too low
values for the parameters could be dangerous. However, high
values could fail when the shocks present in the simulation
were not too strong. 
In order to check conditions (\ref{sh1}-\ref{sh4})
we need a simulation involving shocks, contact discontinuities and 
rarefactions, based on a problem with a well known analytical 
solution. The results of a 3D analytical 
test presented by Quilis et al. (1996) have been used.
The identification of the cells involved in the shock is
excellent in this case. Although general results 
are more complex, the method seems
to be able to identify shocked cells reasonably.
In the practical implementation presented in this paper, 
$\beta$ parameters have been chosen under the constraints that the 
regions labelled in Fig. 6, as shock candidates, are most accurate
recovered.
The values used in this paper for those parameters, which fulfil
the above criteria, are 
$\beta_1=3,\beta_2=2$, and $\beta_3=2$.

Collecting together all the previous prescriptions, 
Figure 7 shows some 3D
plots of the shocked cells at several redshifts for A simulation. 
At high redshifts, the shocked cells form some sort of
filaments which would be following the ones arising in 
DM (see Fig. 1).
Then, we could say that these shocks , which appears at quite
early times, come from the collapse of the substructure and
from merging processes among these substructures. Between 
$z\sim 2$ and $z\sim 0$ the shocked cells mainly trace a 
quasi-spherical shock moving outwards from the cluster center.
Morphologies of the shocks display filamentary structures at 
the early epochs -- high redshift -- while at lower redshift 
a quasi-spherical shock appears at the inner regions and 
propagates outwards dominating the global structure.

%%%%%%%%%%%%%%%%%%%%%%%%%%%%%%%%%%%
%  figure 7, 3d celdas en shocks
%%%%%%%%%%%%%%%%%%%%%%%%%%%%%%%%%%%

The presence of shocks is directly related with the dynamics of gas.
By using the above conditions the number of shocked cells can be
computed. Figure 8 plots the number of shocked cells versus 
time evolution for A(B) continuous(dashed) line.
The maximum number of cells involved in 
shocks is at $z\sim 2$. This epoch is directly correlated with 
those times when the quasi-spherical shock in Figs. 3 and 4 starts to form.

%%%%%%%%%%%%%%%%%%%%%%%%%%%%%%%%%%%%%%%%
% figura 8,grafica con nunera de celdas en shock a lo largo del tiempo
%%%%%%%%%%%%%%%%%%%%%%%%%%%%%%%%%%%%%%%%
  
\subsection{Energy released by shocks}

In previous section the presence of shocks has been discussed.
Now we try to quantify the energy released by those shocks to
ICM.

An easy way to quantify the shock heating is to compare
some quantities obtained by the simulations with the 
values obtained considering an {\it analytical adiabatic evolution}.
The quantities derived from this  analytical adiabatic evolution
are labelled with  tilde (\,$\tilde{}$\,).
As it is well known in the adiabatic regime the pressure and 
temperature are
related with density at any time in the following form:
\be
\label{ad1}
\tilde{p}=\kappa \tilde\rho^{\gamma}_b
\en
\be
\label{ad2}
\tilde{T}=\eta \tilde\rho^{\gamma-1}_b
\en
where $\kappa$ and $\eta$ are two constants. 
In order to fix the values of these constants we take the same values of 
pressure (p), density ($\rho$), and temperature (T) as the ones used, at 
the initial time, in the numerical simulations.

If the simulated cluster would evolve pure adiabatically,  
Eqs. (\ref{ad1}-\ref{ad2}) should give, at any time, 
the same values of pressure and temperature than those obtained
from the hydro-code. In particular, if the 
baryonic density derived from  
the code were used as an  input for the above adiabatic laws,
and the cluster would evolve pure adiabatically, 
no differences should be found, at any time, 
between results from Eqs. (\ref{ad1}-\ref{ad2})
and the ones given by the numerical code.
In practice, we compute all the pure adiabatic quantities
($\tilde{p},\tilde{T},\tilde{\rho}_b$) using as input the density 
coming from the simulations. 
Any deviations between results from the simulation A, and results
from Eqs. (\ref{ad1}-\ref{ad2}) --  where it has been used the density
evolved by the A simulation -- should be considered as a proof 
of the presence of non-adiabatic processes.

The total internal energy for the whole box is defined as :
\be
E_U=\int_V \rho_b \epsilon dV
\en
\noindent
being $V$ the physical volume.

Three total internal energies can be computed. First(second) 
is the one derived from the A(B) simulation. A third one is obtained 
from the
analytical adiabatic evolution by using Eqs. (\ref{ad1}-\ref{ad2}).
Top panel of  Figure 9 shows
total internal energy from A simulation (continuous line)
and  B simulation (dashed-pointed line), 
and from analytical adiabatic evolution 
(dashed line).
The curves from simulations appear clearly different
from the analytical case.
If the evolution were completely adiabatic, the internal
energy for the A simulation (no cooling effects) should be
identical to the one for the analytical adiabatic evolution.
As this simulation does not consider cooling effects, all these
differences are due to non-adiabatic processes.
That means that shocks are the responsible of the most part of the ICM 
heating.
The total internal energy for the B simulation is slightly lower than 
the one obtained in the A case  at some phases of the evolution. 
These small differences are negligible in front of the differences
between A or B cases against the analytical adiabatic evolution curve. 
Consequently, cooling effects are negligible compared with shocks.

Total kinetic energy is estimated from the following volume integral: 
\be
E_K=\frac{1}{2}\int_V \rho_b v^2 dV
\en

Bottom panel of Figure 9 plots time evolution of the
ratio $E_U/E_K$ for the data generated by our simulations. 
Continuous (dashed-pointed) line corresponds to A(B) simulation.
These curves show those times of the evolution where the internal energy
is being released to the ICM by shocks. 
The decreasing branch in Figure 8, after the first maximum, is
related with the epoch in which shocks -- already formed -- 
are propagating and loosing
energy, which contributes to the ICM heating. This branch is
correlated with the raising branch of the $E_U/E_K$ curve in Figure 9.

%%%%%%%%%%%%%%%%%%%%%%%%%%%%%%%%%%%%%%%%%%%%%%
%  figura 10, energia intera adiabarica y cinetica
%%%%%%%%%%%%%%%%%%%%%%%%%%%%%%%%%%%%%%%%%%%%%%

Figure 10 is a plot of the time evolution of the logarithm of the average 
temperature in a ball of one comoving Mpc radius centered at the core cluster. 
As usual, data generated by A (continuous line) and B (dashed-pointed line) 
simulations, and the analytical adiabatic evolution (dashed line) are 
displayed. 
As the observations suggest, the values of temperature obtained in
our simulations range around $\sim 3\times 10^7 \, K$ at low redshifts.
Both simulations, A and B, show how the temperature has increased 
about six orders of magnitude over that of
the pure adiabatic evolution.  There are
no important differences between A and B temperatures,
as in the case of the quantities analyzed previously.

Figure 11 plots -- for the case B -- the X-Ray thermal bremsstrahlung 
luminosity produced by 
a ball of one comoving Mpc radius centered at the core cluster.
The values are compatible with those given by the X-Ray 
observations which range between $10^{42}$ and $10^{44}$ erg/s. 
 
%%%%%%%%%%%%%%%%%%%%%%%%%%%%%%%%%%%%%%%%%
% Figura 11 , temperatura y luminosidad 
%%%%%%%%%%%%%%%%%%%%%%%%%%%%%%%%%%%%%%%%%

\section{Discussion}

In this paper we have used some numerical techniques 
recently applied to Cosmology. These techniques 
, HRSC, seems to be the most suitable in order
to study the role of shocks
in galaxy cluster evolution. 
The choice  
is justified by their properties to handle shocks.
The capability of these techniques to capture shocks 
with very small diffusion is independent of the resolution 
used in the numerical simulations. Hence, by construction,
shocks are captured even using coarse grids. This property
is crucial in 3D applications.

Previous sections illustrate the fact that non adiabatic
processes, due to shocks,
 take an important role in the description of
the ICM. In the model presented in this paper,
that is, a baryonic fluid plus
dark matter component coupled gravitationally, shocks 
are able to heat the ICM until values compatible with
observational data.

The calculations have been carried out in two cases: 
with and without cooling processes. This procedure
allows us to distinguish between non-adiabatic
effects coming from shocks and the ones from cooling. The role of 
the cooling, even when 
it could be important in other scenarios, 
is  irrelevant for the simulations considered in this paper, 
while shocks play the most important role.

In the picture describing the dynamics of the baryonic
component, there are some clues showing the presence of shocks. 
Examining  the quantities sensitive to shocks, all of them evidence
the formation of a quasi-spherical shock.
This shock seems to arise around $z\sim 2$ 
at the cluster center and
moves outwards. 
Nevertheless, some irregular shocks could form at $z \geq 2$. 
This conclusion is supported by the behaviour of the entropy profiles 
(see Fig. 5), and
the existence of shocked cells at these times (see Fig. 8).
The quasi-spherical shock 
would form from the collapse of the quasi-spherical global structure , 
while other smaller shocks -- with a filamentary morphology --
would arise from some collapsing substructure and merging processes.
In short, previous discussion manifest two different regimes in the 
shock formation.

It should be noticed that the structures simulated in this paper
correspond, due to the initial conditions, to a large Abell cluster.
For this kind of clusters, gravitational collapse is fast and 
the dynamics is violent. Shocks form earlier and are stronger 
than in  others smaller cluster-like objects ($ < 3\sigma$).

Some discussion on the numerical resolution of the simulations is needed.
The one used in present paper ($\sim 0.3 Mpc$) is 
not enough to simulate the very center of the clusters and 
galaxy formation, but 
it suffices to study the role of the shocks in ICM.
It should be kept in mind that HRSC techniques are able
to resolve shocks even with coarse grids. Nevertheless, higher resolution
would be desiderable to perform more complete simulations.
Improvements in numerical resolution will introduce smaller scales
in the problem, as a consequence, the physics of the model 
should be enriched in order to describe this new scenario.
Chemical reactions and  radiative 
transfer should be considered.  
  
%%%%%%%%%%%%%%%%%%%%%%%%%%%%%%%%%%%%%%%%%%%%%%%%%%

\acknowledgements 
This work has been 
supported by the Spanish DGES(grants PB96-0797 and PB94-0973).
V. Quilis thanks to the Conselleria d'Educaci\'o i 
Ci\`encia de la Generalitat Valenciana for a fellowship.
Dr. P. Anninos and Dr. J.M$^{\underline{\mbox{a}}}$. Mart\'{\i} are
 acknowledged for useful discussions.
One of the authors, V.Q., thanks 
Dr. S.M.D. White for the hospitality
during 
his visit to {\it Max-Planck-Institut f\"ur Astrophysik}.
Authors gratefully acknowledge the enlightening comments of
the referee.
Calculations were carried out in a
HP J280 and in two SGI Origin 2000, at the {\it Centre 
Europeu de Paral$\cdot$lelisme de Barcelona (CEPBA)}
and the {\it Centre de Inform\'{a}tica de la Universitat de
Val\'{e}ncia}.

\appendix
\section{Energy conservation properties}
As it is well known,
in  simulations without cooling the total energy  
must be conserved. This conservation law can be
obtained after integrating on the whole computational volume
the evolution equations, and 
it reads as follows (Peebles 1980):
\begin{equation}
\label{conE1}
\frac{d(\tilde{E}+W)}{dt}+\frac{\dot{a}}{a}(2\tilde{E}+W)=0
\end{equation}
\noindent
where $\tilde{E}$ is the total energy -- kinetic of the gas plus kinetic of
the particles plus internal for the gas--  and $W$ is the total
gravitational energy, i.e., gas plus particles.

The Eq.(\ref{conE1}) can be intregated respect to the scale factor 
giving:
\begin{equation}
\label{conE2}
a(t_2)\tilde{E}(t_2)-a(t_1)\tilde{E}(t_1)
+\int_{a(t_1)}^{a(t_2)}\tilde{E} da 
= -(a(t_2)W(t_2))-a(t_1)W(t_1))
\end{equation}
\noindent
being $t_2$ and $t_1$ two
different times. The energy conservation can be tested by defining
a quantity, $R$, as:
\begin{equation}
\label{conE3}
R(t)=\frac{a(t)\tilde{E}(t)-a(t_1)\tilde{E}(t_1)
+\int_{a(t_1)}^{a(t)}\tilde{E} da}
{-(a(t)W(t))-a(t_1)W(t_1))}
\end{equation}
\noindent
Consequently with Eq. (\ref{conE2}), quantity $R$ must remain equal 
to unit during the evolution.

We have tested the energy conservation for 
the whole code
, i.e., hydrodynamical and particle codes, coupled through
the multidimensional Poisson solver. One numerical 
simulation  with  
the same initial condition as in the case A, and with  
$64^3$ cells and $64^3$ particles has been considered. 
Figure 12 plots the quantity $R$ against redshift. The maxima
errors are around three per cent. As it is well known, 
the better resolution (finer grids and more particles), the
better conservation properties.

\section{Some comments on the resolution of the hydrodynamical code}

Due to the coarse grids we have used in this paper, our
results depend strongly on the spatial resolution.
Our numerical simulations  have been performed 
using the hydro-code described in Quilis et al. (1996).
In that paper the authors discussed two types of reconstruction
procedures as methods to increase the spatial resolution of 
the algorithm, i.e. MUSCL (linear) and PPM (parabolic). 
In order to investigate the influence of the numerical 
resolution in our galaxy cluster simulations, 
we have considered some numerical experiments.
In order to do that,
some simulations without cooling and with the same initial 
conditions than in A case (see Section 2) have been performed.

The above two reconstruction procedures have been considered.
Together with the reconstruction, we have studied the influence of the 
number of cells and particles used in the simulations. 
Extremely coarse grids with $64^3$ cells and $64^3$ particles,
and the one used in the simulations in this paper, 
$128^3$ cells and $128^3$ particles, have been considered.

Table 1 summarizes the main results of these experiments.
We show the maxima of the density contrast of the gas 
 , $\delta_b^{max}$, at two
times. At each time, these maxima are presented for both
grids and set of particles , i.e. $64^3$ and $128^3$, and for 
MUSCL and PPM reconstructions. 
As it follows from that table, with the grids considered, 
MUSCL does not work properly.
PPM appears as the best
procedure to increase the spatial resolution, in this sense,
the simulations presented in this paper have been computed 
using PPM reconstruction procedure. 

\newpage

\begin{table}
\begin{center}
\caption{Influence of the numerical resolution.}
\begin{tabular}{cccccc}
         & z & MUSCL64 & PPM64 & MUSCL128 & PPM128 \\
$\delta_b^{max}$ & 1 &   24.5  &  56.3 &   53.2   &  210.5 \\
$\delta_b^{max}$ & 0 &   11.5  &  96.0 &   31.5   &  348.8 \\
\end{tabular}
\end{center}
\end{table}

\newpage
 
%Figure Captions
 
\figcaption{Dark matter particles distribution inside the computational
boxes for six epochs for A simulation.
Boxes are 20 comoving Mpc length per edge. Only 32$^3$ particles are
displayed in each box.} 
 
\figcaption{Slices along z-axis of the gas density contrast 
for A simulation. Each slice is 
20 comoving Mpc length per edge and 0.15 comoving Mpc depth.
Columns stand for fixes redshift. Central rows are centered at
the maximum density contrast, up(low) row is located at -5(+5)
comoving Mpc from the central slices along z-axis. At left side
of each slice there is one palette describing the colour scale 
used to plot it.}

\figcaption {Average density contrast, for A simulation, 
in radial shells for gas(top) and DM(bottom). 
Six times are displayed. Radii are
in comoving Mpc.}

\figcaption {Average radial velocities, for A simulations, 
in radial shells for 
gas(top) and DM(bottom). 
Six times are displayed. Radii are
in comoving Mpc and velocities in units of speed of light
($c$).}

\figcaption {Average entropy in radial shells for 
gas in A simulations.
Six times are displayed. Radii are
in comoving Mpc. Entropy is defined as $s=ln(T/\rho_{_b}^{\gamma -1})$.}

\figcaption{From top to bottom, slices show (for the baryonic
component): divergence of  
velocity in units of c(first row), entropy (second row), 
temperature in K (third row), and
pressure in $dyn/cm^2$(fourth row).
Each slice is 
20 comoving Mpc length per edge and 0.15 comoving Mpc depth.
Columns stand for a fixed redshift. All slices are centered at
the maxima of the density contrast.
At left side
of each slice there is one palette describing the colour scale. 
Results correspond to A simulation.}

\figcaption{Shocked cells  inside the computational
boxes for six times for A simulation.
Boxes are 20 comoving Mpc length per edge.} 
 
\figcaption{Number of shocked cells ($N_{shc}$) 
versus redshift. 
Continuous(dashed) line plots A(B) simulation.}

\figcaption{Plot of logarithm of the total internal energy ($E_U$) 
against redshift (top
panel) and the ratio of the total internal energy ($E_U$) to the total 
kinetic ($E_K$) energy versus
redshift (bottom panel). All energies are in units of ergs.
Continuous (dashed-pointed) line 
stands for A(B) results,
in top panel dashed line shows analytical adiabatic result.} 

\figcaption{Plot of 
the logarithm of the average temperature (in K degrees)
in a ball of 1 comoving Mpc  centered at the core cluster.
Continuous, dashed-pointed, dashed  lines 
stands for A and B simulations, and 
analytical adiabatic evolution, respectively.} 

\figcaption{Plot for B simulation  of 
the logarithm of the X-Ray luminosity (in $erg/s$),  from
a ball of 1 comoving Mpc,  centered at the core cluster.}

\figcaption{Plot for the quantity R defined in 
Eq. (\ref{conE3}) against the redshift. Perfect energy
conservation would imply $R=1$.}


\newpage

\begin{thebibliography}{}

\bibitem[]{}Anninos, P., Norman, M.L. \& Clarke, D.A. 
1994, ApJ, 436, 11

\bibitem[]{}Anninos, P. \& Norman, M.L. 1996, ApJ, 459, 12


\bibitem[]{}Bertschinger, E. \& Gelb, J.M. 1991, Computer in Physics,
 5, 164

\bibitem[]{}Bryan, G.L., Cen, R.,Norman, M.L.,Ostriker, J.P.
\& Stone, J.M. 1994, ApJ, 428, 405

\bibitem[]{}Cen, R. 1992, ApJS, 78, 341

\bibitem[]{}Colella, P. \& Woodward, P.R. 1984, J. Comp. Physc,
54, 174

\bibitem[]{}Einfeldt, B., Munz, C.D., Roe, P.L. \& Sj\"ogreen, B.
1991, J. Comp. Phys., 92, 273

\bibitem[]{}Evrard, A.E. 1988, MNRAS, 235,911


\bibitem[]{}Efstathiou, G., Davis, M., Frenk, C.S. \&
White, S.M.D. 1985, ApJS, 57, 241


\bibitem[]{}Gingold, R.A. \& Monaghan, J.J. 1977, MNRAS, 181, 375

\bibitem[]{}Gnedin, N.Y. 1995, ApJS, 97, 231

\bibitem[]{}Godunov, S.K. 1959, Matematicheskii Sbornik, 47, 271

\bibitem[]{}Guzzo, L. 1996, in ASP Conference Series 94,
Mapping, Measuring, and Modelling the Universe, ed. Coles, P.,
Martinez, V.J., and Pons-Borderia M.J. (San Francisco, 
Astronomical Society of the Pacific) 157

\bibitem[]{}Harten, A. 1983, J. Comp. Phys., 49, 357

\bibitem[]{}Hernquist, L. \& Katz, N. 1989, ApJS, 64,715

\bibitem[]{}Hockney,R.W. \& Eastwood, J.W. 1988, Computer 
simulation usign particles. IOP Publishing

\bibitem[]{}Hoffman, Y. \& Ribak, E. 1991, ApJ, 380, L5


\bibitem[]{}Kang, H., Ostriker, J.P., Cen, R., Ryu, D.,
Hernquist, L., Evrard, A.E., Bryan, G.L. \& Norman, M.L.
 1994, ApJ, 430, 83

\bibitem[]{}Lucy, L.B. 1997, AJ, 82, 1013

\bibitem[]{}Navarro, J.F. \& White, S.D.M. 1993, 
MNRAS,265, 271

\bibitem[]{}Navarro, J.F., Frenk, C.S. \& White, S.D.M. 1995, 
MNRAS,275, 720
 
\bibitem[]{}Peebles, P.J.E. 1980, The Large-Scale Structure of 
the Universe (Princeton: Princeton Univ. Press) 
 
\bibitem[]{}Press, H., Flannery, B.P., Teukolsky, S.A. \&
Vetterling, W.T., 1987,  Numerical Recipies. The art of
Scientific Computing. Cambridge University Press

\bibitem[]{}Quilis, V., Ib\'a\~nez, J.M$^{\underline{\mbox{a}}}$. 
\& S\'aez, D. 1996, ApJ, 469, 11

\bibitem[]{}Roe, P.L. 1981, J. Comp. Phys., 43, 357

\bibitem[]{}Ryu, D., Ostriker, J.P., Kang, H. \& Cen, R.
1993, 414, 1 

\bibitem[]{}Tsai, J.C., Katz, N. \& Bertschinger, E. 1994, 423, 553

\bibitem[]{}Umemura, M. \& Ikeuchi, S. 1984, Prog. Theor. Phys. 72,47

\bibitem[]{}van den Weygaert, R. \& Bertschinger, E. 1996, MNRAS, 281, 84

\bibitem[]{}Xu, G. 1995, ApJS, 98, 355

\end{thebibliography}
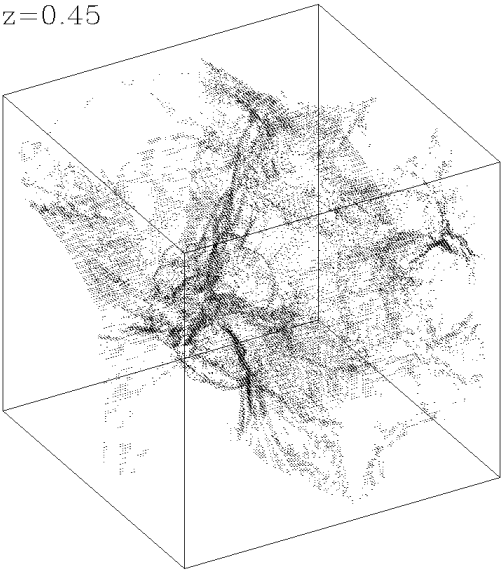
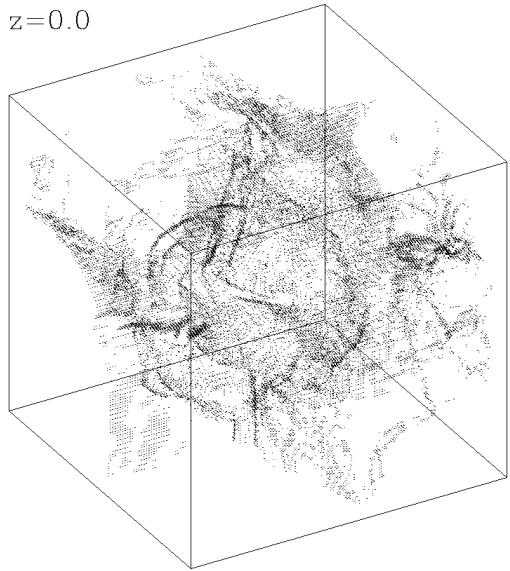
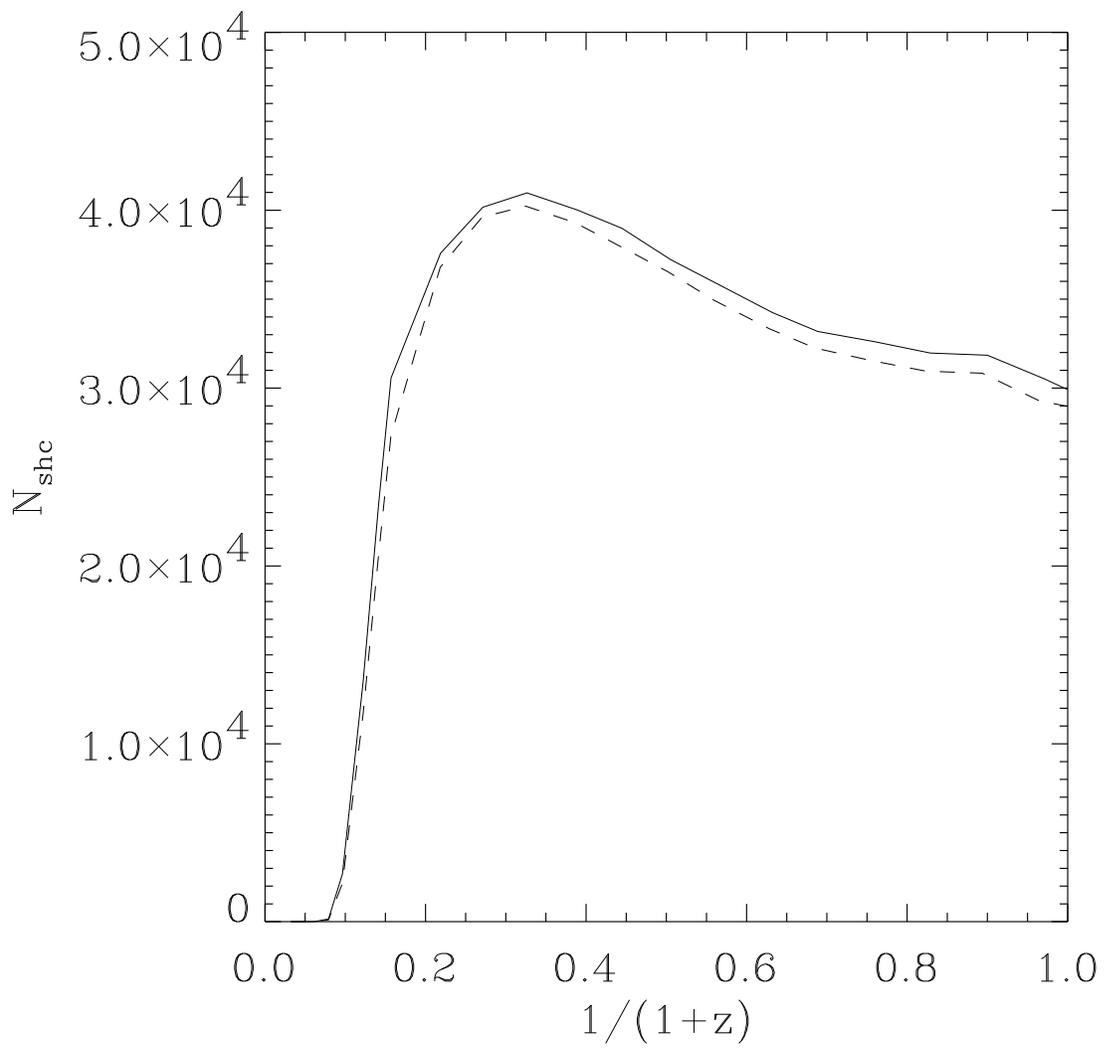
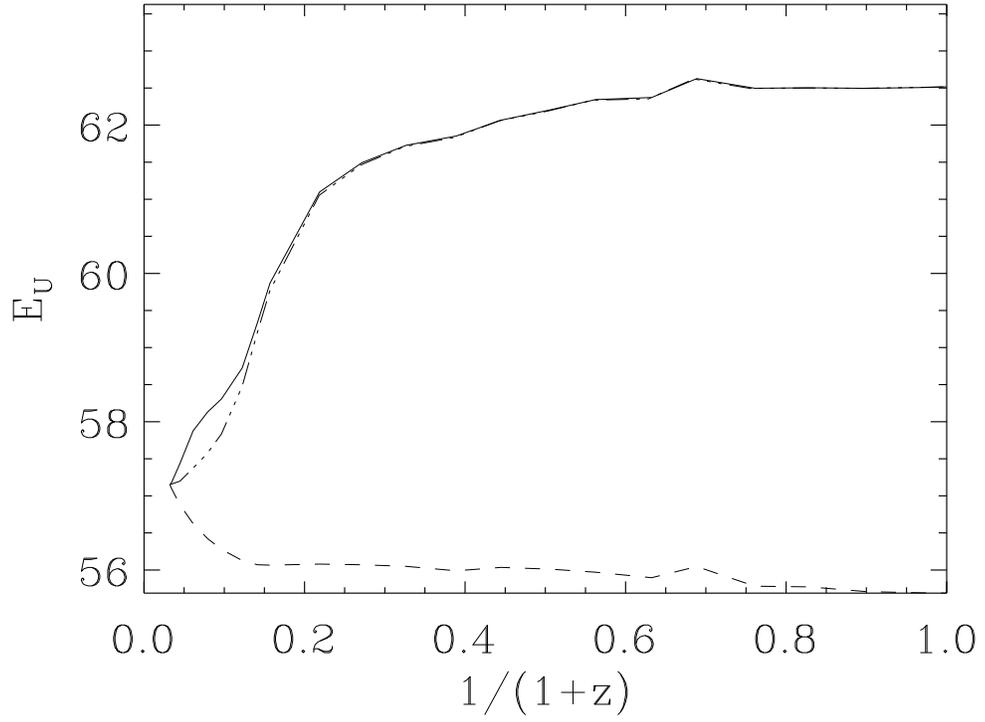
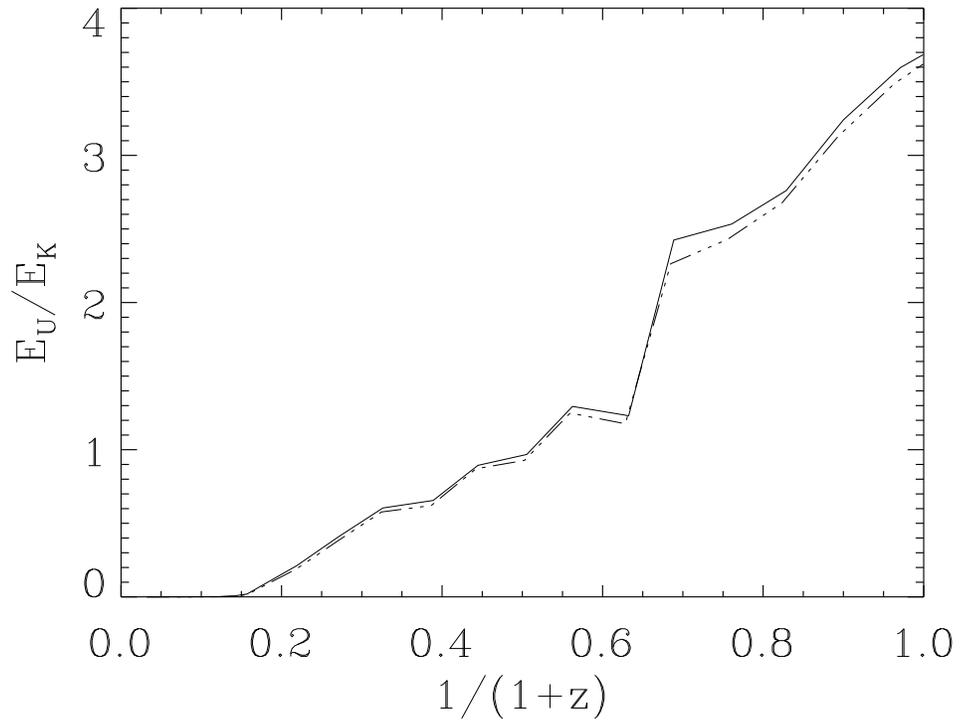
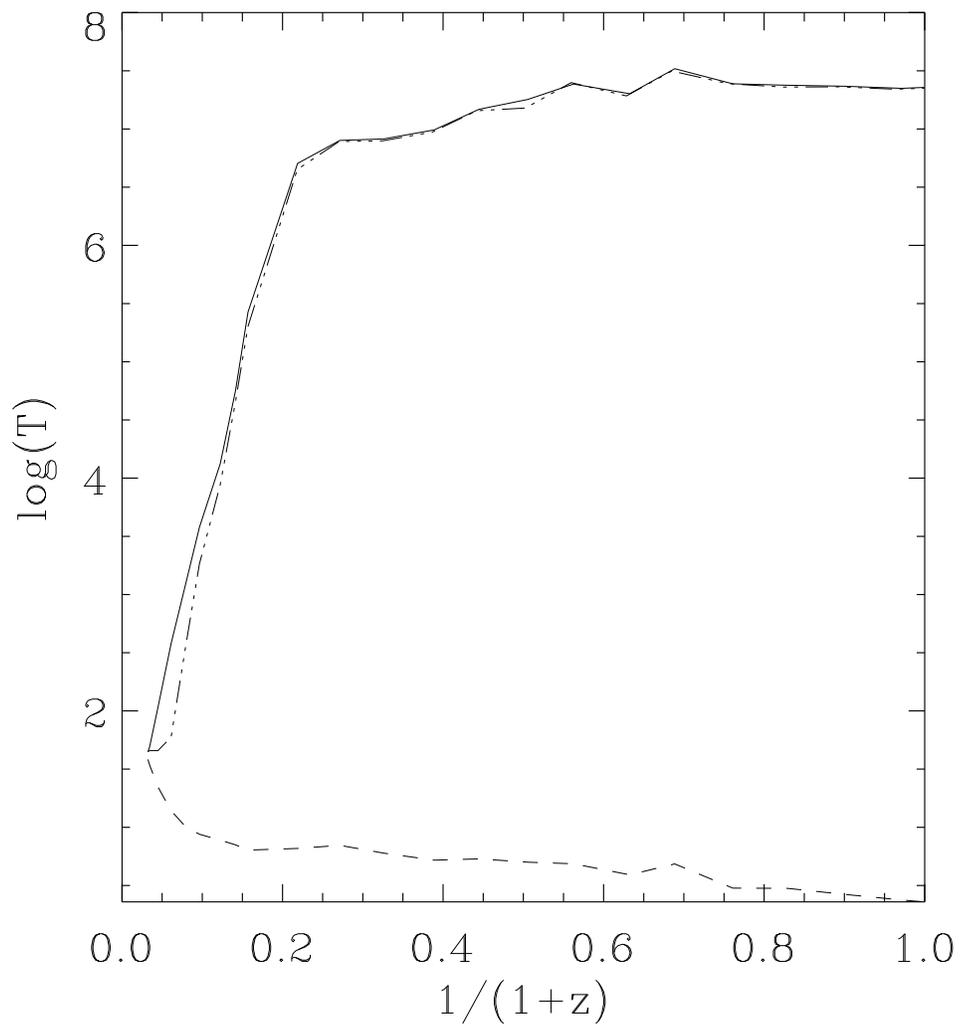
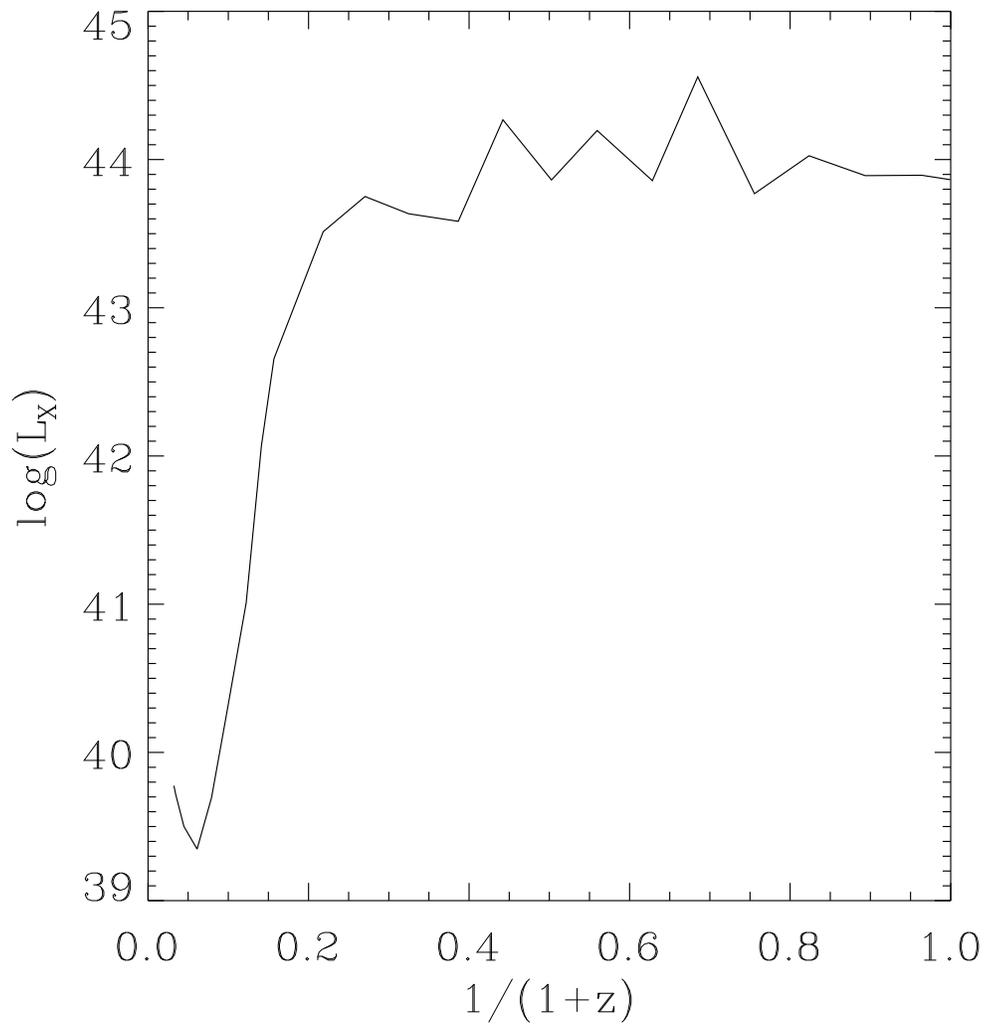
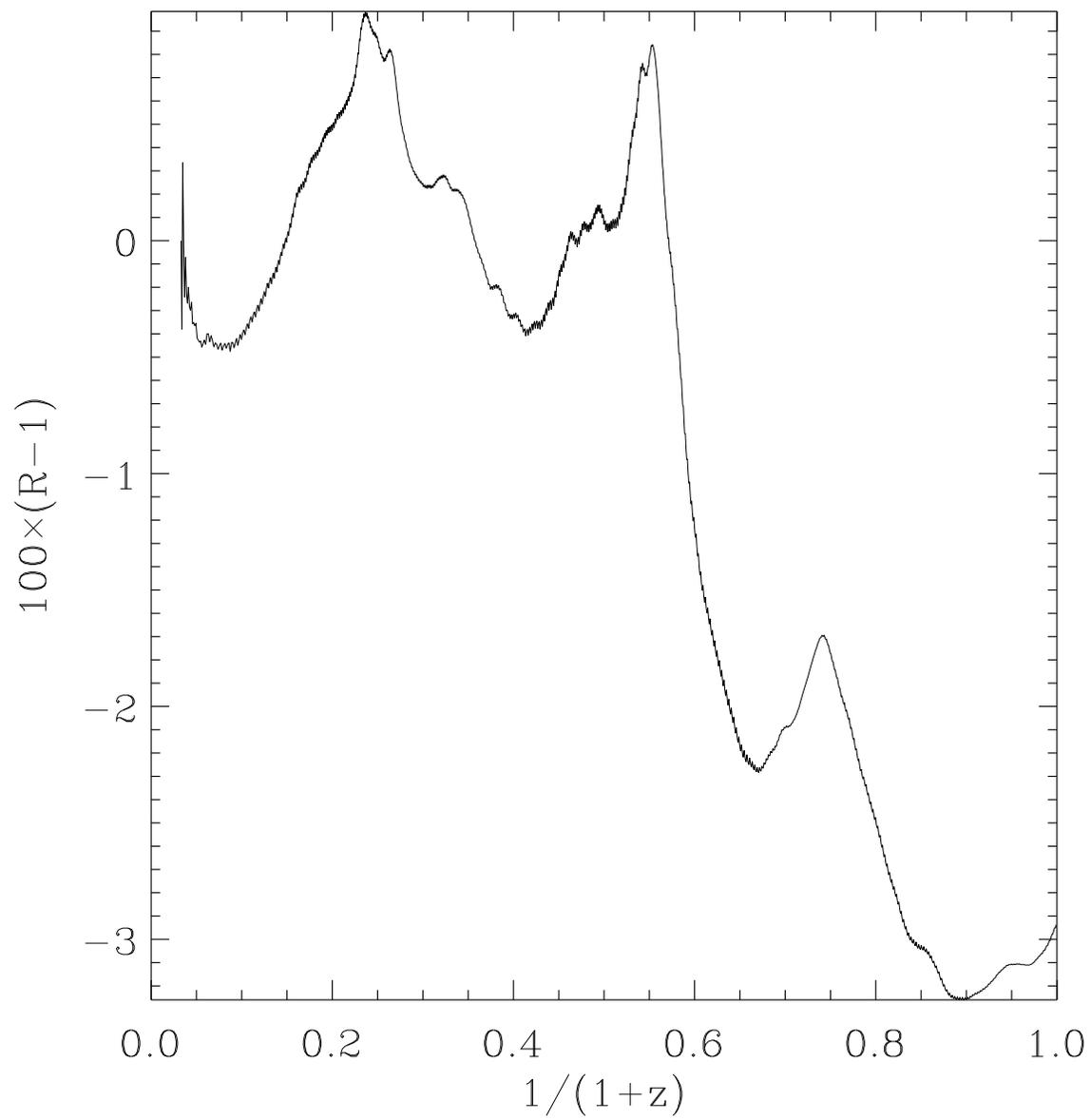
\end{document}